\documentclass{appolb}
\usepackage{graphicx}

\begin{document}
\title{QCD Critical Point and High Baryon Density Matter %
\thanks{Presented at workshop on ''Criticality in QCD and the Hadron
  Resonance Gas'', Wroclaw (online), July 29-31, 2020}%
}
\author{B. Mohanty$^{1,2}$ and N. Xu$^{2,3}$
\address{
$^1$School of Physical Sciences, National Institute of Science
  Education and Research, HBNI, Jatni 752050, India, $^2$Institute of Modern Physics, 
  509 Nanchang Road, Lanzhou 730000, China and $^3$Nuclear Science
  Division, Lawrence Berkeley National Laboratory, Berkeley, CA 94720,
  USA}
}
\maketitle
\begin{abstract}
We report the latest results on the search for the QCD critical point
in the QCD phase diagram through high energy heavy-ion collisions. The
measurements discussed are based on the higher moments of the net-proton
multiplicity distributions in heavy-ion collisions. A non-monotonic
variation in the product of kurtosis times the variance of the net-proton
distribution  is observed as a function of the collision
energy with 3$\sigma$ significance. We also discuss the results of the
thermal model in explaining the measured particle yield ratios in
heavy-ion collisions and comparison of the different variants of hardon
resonance gas model calculation to the data on higher moments of net-proton
distributions. We end with a note that the upcoming programs in high
baryon density regime at various experimental facilities will complete
the search for the QCD critical point through heavy-ion collisions.
\end{abstract}
\PACS{25.75.-q,25.75.Nq, 12.38.Mh, 12.38.-t,25.75.Gz}
  
\section{Introduction}

\begin{figure}[htb]
\centerline{%
\includegraphics[width=10.0cm]{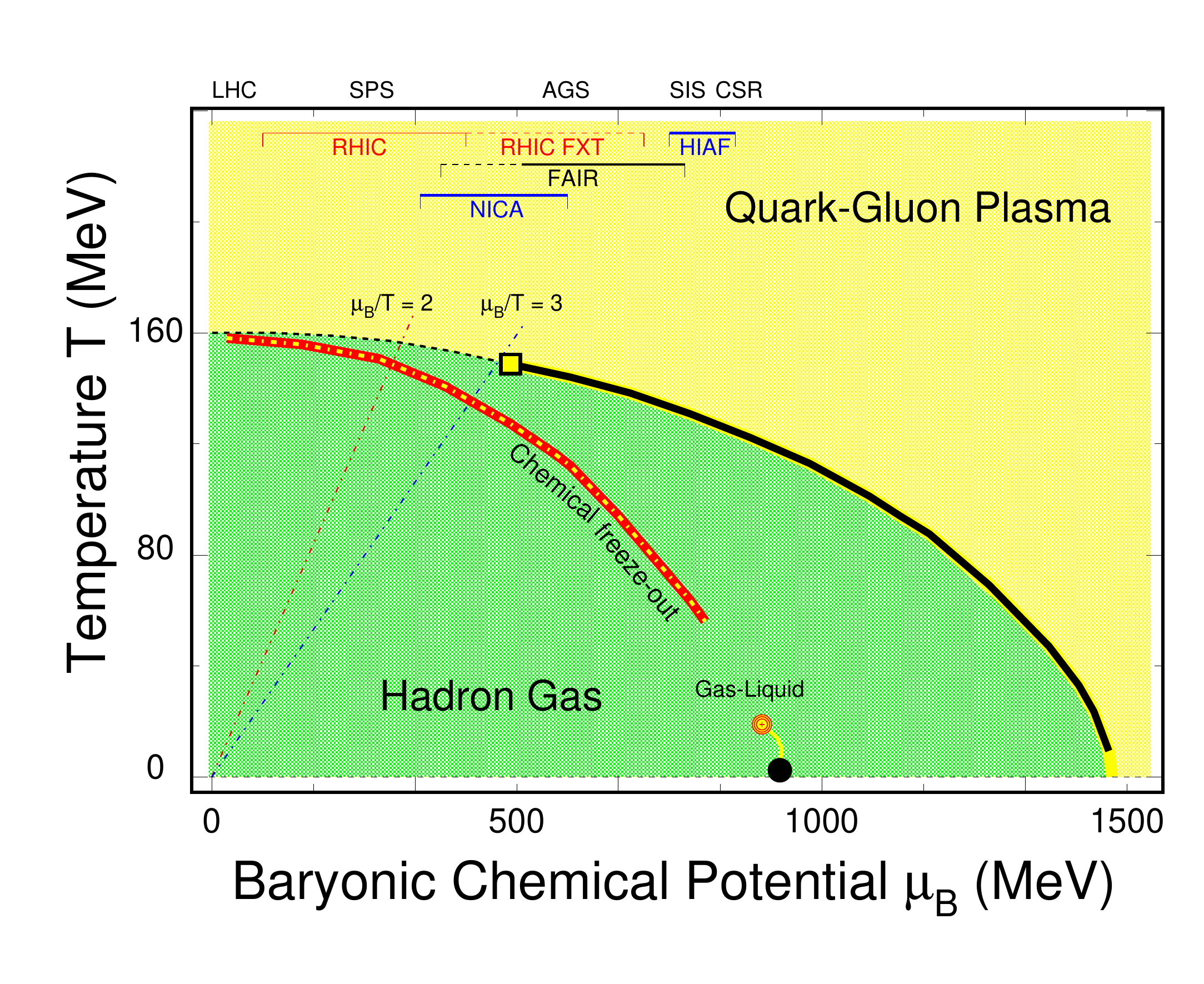}}
\caption{Conjectured QCD phase diagram of temperature ($T$) versus baryonic 
  chemical potential ($\mu_{\mathrm B}$). See text for details.}
\label{Fig:F1}
\end{figure}

Relativistic heavy-ion collisions at varying center of mass energy ($\sqrt{s_{NN}}$)
allows for the study of the phase diagram of nuclear matter~\cite{starnote}. The
underlying theory is the one that governs the strong interactions -
Quantum Chromodynamics (QCD). The conjectured phase diagram of QCD is
shown in Fig.~\ref{Fig:F1}. The current status of the phase diagram is as
follows. There are two distinct phases in the phase structure:
de-confined state of quarks and gluons called the quark gluon plasma
(QGP) and the confined state of gas of hadrons and resonances (HRG).
The phase boundary (shown as a solid line in Fig.~\ref{Fig:F1}) between  the hadronic gas phase and 
the high-temperature quark-gluon phase is a first-order phase
transition line, which begins at large baryon chemical potential
($\mu_{B}$) and small temperature ($T$)
and curves towards smaller $\mu_{B}$  and larger $T$. This line ends 
at the QCD critical point whose conjectured position, indicated by a square, is uncertain both 
theoretically and experimentally. At smaller $\mu_{B}$ there is a cross over indicated by a dashed line.
The region of $\mu_{\mathrm B}$/$T$ $\leq$ 2 is shown as 
dot-dashed line. A comparison between RHIC data and lattice QCD (LQCD)
calculations disfavours
the possible QCD critical point being located at $\mu_{\mathrm
  B}$/$T$ $\leq$ 2~\cite{Bazavov:2017tot,Bazavov2:2017dus}.
The red-yellow dotted line corresponds to the
chemical freeze-out obtained from the fits of particle yields in
heavy-ion collisions using a thermal model.
The liquid-gas transition region features a second order critical point (red-circle) and a
first-order transition line (yellow line) that connect the critical point to
the ground state of nuclear matter ($T$ $\sim$ 0 and $\mu_{\mathrm B}$ $\sim$ 925
MeV)~\cite{Fukushima:2010bq}.
The regions of the 
phase diagram accessed by past (AGS and SPS), ongoing (LHC, RHIC, SPS
and  RHIC operating in fixed target mode), and future (FAIR and 
NICA) experimental facilities are also indicated.

In this proceeding, we discuss the success and tests of the hadron
resonance gas model using the particle ratios and fluctuations in
net-proton number produced in heavy-ion collisions. We also discuss
the status of the search for the QCD critical point and future
experimental directions in this connection at the upcoming facilities.

\section{Particle ratio and thermal model}

Thermal models, assuming approximate local thermal equilibrium, have
been successfully applied to matter produced in heavy-ion
collisions. Most popular variant of such a model employs Grand
Canonical Ensemble (GCE), hence uses chemical potentials to account for
conservation of quantum numbers on an
average~\cite{Andronic:2017pug}. For systems created via elementary
collisions (small system) or via low energy heavy-ion collisions, the Canonical Ensemble
(CE) approach is used. In the large volume limit, the GCE and the CE
formalisms should be equivalent. In heavy-ion
collisions at energies spanning from few GeV to few TeV it may be
worthwhile to ask at what collision energy a transition from GCE to CE
occurs~\cite{Hagedorn:1984uy} ?

\subsection{Success of thermal model}
\begin{figure}[htb]
\centerline{%
\includegraphics[width=10.0cm]{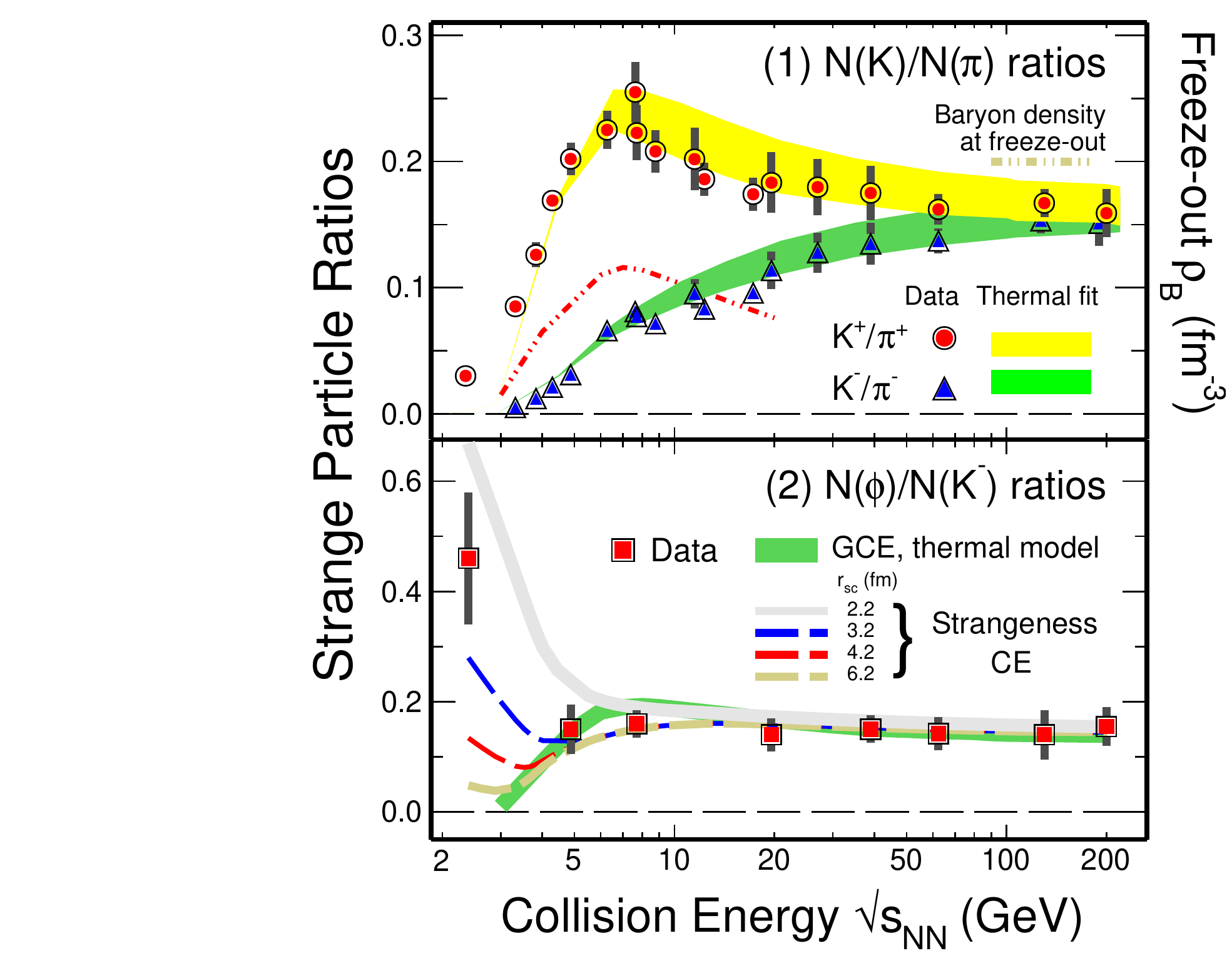}}
\caption{(1) Ratio of yields of kaon to pion ($K^+/\pi^+$
    (circles) and  $K^-/\pi^-$ (triangles) produced in central heavy-ion 
  collisions at mid-rapidity as a function of $\sqrt{s_{NN}}$. Thermal 
    fits are also shown as bands (yellow band for $K^{+}$/$\pi^{+}$
    and green band for $K^{-}$/$\pi^{-}$) in the plot. Dot-dashed line represents 
    the net-baryon density at the chemical freeze-out.  The 
dot-dashed line represents the net-baryon density at the Chemical 
Freeze-out as a function of collision energy, calculated from the 
thermal model~\cite{Randrup:2006nr}. 
 (2) Ratio of 
    yields of $\phi$-meson to kaon ($\phi/K^-$) produced in central heavy-ion collisions at 
    mid-rapidity as a function of $\sqrt{s_{NN}}$. The various bands 
    shows the thermal model expectation from grand canonical ensemble 
    (GCE) and canonical ensemble (CE) formulations in the HRG model.}
\label{Fig:F2}
\end{figure}

Figure~\ref{Fig:F2}~(1) in the upper panel shows the energy 
dependence of $K$/$\pi$ particle yield ratio produced in heavy-ion
collisions at AGS~\cite{Ahle:1999uy,Ahle:1999va,Ahle:2000wq},
SPS~\cite{Afanasiev:2002mx,Alt:2007aa} and
RHIC~\cite{Adamczyk:2017iwn}.  The thermal model calculation explains 
the   $K$/$\pi$ ratios that reflect the strangeness 
content relative to entropy of  the system formed in heavy-ion 
collisions. This can be treated as a success of the application of
thermal model to heavy-ion collisions. 
A peak in the energy dependence of $K^{+}$/$\pi^{+}$ could be due to associated production
dominance  at lower energies as the baryon stopping is large.  The
peak is consistent with the calculated net baryon
density reaching a maximum~\cite{Randrup:2006nr} has been suggested to be a signature 
of a change in  degrees of freedom (baryon to 
meson~\cite{Cleymans:2004hj} or hadrons to 
QGP~\cite{Gazdzicki:1998vd}) while going from lower to higher 
energies. The  $K^{-}$/$\pi^{-}$ ratio seems unaffected by the changes in the 
net-baryon density with collision energy and shows a smooth increasing 
trend.

\subsection{Transition from grand canonical to canonical ensemble}

Figure~\ref{Fig:F2}~(2) in the lower panel shows the energy 
dependence of $\phi$/$K^{-}$ yield ratio measured in heavy-ion
collisions~\cite{Abelev:2008zk,Adam:2019koz,Agakishiev:2009ar}. As one moves from higher to lower collision energy, the 
$\phi$/$K^{-}$ ratio changes rapidly from a constant value to larger
values. The transition happens below the collision energy where the
freeze-out net-baryon density peaks (see upper panel). Thermal model
calculations with GCE explains the measurements up to collision energy
of $5$ GeV.  At lower energies the GCE model expectation is that the
$\phi$/$K^{-}$ ratio should decrease in contrast to that observed in
experiments. On the other hand, the increase in $\phi$/$K^{-}$ at lower 
energies is explained by thermal model with  CE framework for
strangeness production. 
The results are also sensitive to the choice of the additional control
parameter, $r_{\rm sc}$, in CE framework, which decides the typical
spatial size of $s\bar{s}$ correlations. Hence, we find that a high
statistics and systematic measurement of $\phi$/$K^{-}$ yield ratio
can be used to test the transition of GCE to CE in thermal models. As
the size of the $s\bar{s}$ correlations depends on the medium
properties, such studies will provide valuable data for estimation of the
volume in which open strangeness is produced.

\section{Net-proton number fluctuations and QCD critical point}

The QCD critical point is a landmark on the QCD phase diagram. 
Experimental signatures for critical point is enhanced
fluctuations coupled to the critical modes. In this respect the baryon
number fluctuations are sensitive to the criticality~\cite{Hatta:2003wn}. At the critical
point, generally, the correlation length takes large values, and
that leads to non-Gaussian fluctuations~\cite{Stephanov:2008qz}. 
Higher-order fluctuations are more sensitive to the criticality, the third order ($S\sigma$)
and the fourth order ($\kappa\sigma^2$) are common
measures for the QCD critical point search, where
$\sigma$, $S$ and $\kappa$ are called the standard deviation, skewness
and the kurtosis of the distribution, respectively.  Experimentally,
net-proton distribution is considered as a proxy for net-baryon distributions.

\subsection{Net-proton number fluctuations}
\begin{figure}[htb]
\centerline{%
\includegraphics[width=11.5cm]{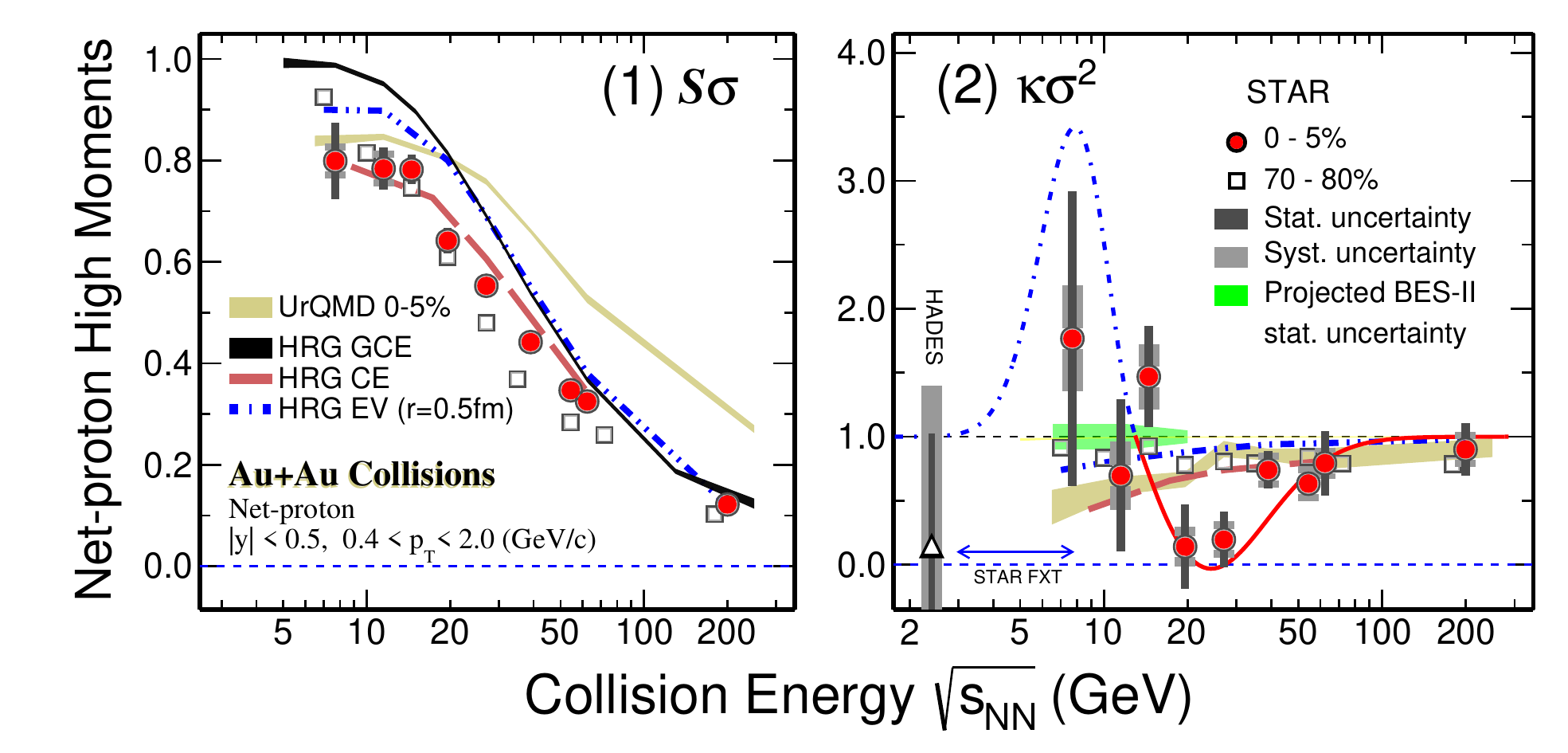}}
\caption{(1) $S\sigma$ and (2)  $\kappa\sigma^2$ of net-proton
  distributions for 70-80\% peripheral  (open squares) and 0-5\% central  (filled-circles) Au+Au 
    collisions as a function of $\sqrt{s_{NN}}$~\cite{Adam:2020unf}. Projected statistical uncertainty for the second phase 
    of the RHIC BES program is shown by the 
    green-band and the blue arrow shows the region of  $\sqrt{s_{NN}}$
    to be covered by the STAR experiments fixed-target program. Results of calculations 
    are shown for different variants (Ideal GCE \cite{Garg:2013ata}, excluded volume~\cite{Bhattacharyya:2013oya} and
    CE~\cite{Braun-Munzinger:2020jbk}) of HRG model and transport 
    model (UrQMD). The solid red and the dashed  blue line in (2) is a
    schematic representation of expectation from a  QCD based model
    calculation in presence of a critical point.}
\label{Fig:F3}
\end{figure}

Figure~\ref{Fig:F3} shows the most relevant measurements over the
widest range in $\mu_{B}$ ($20-450$ MeV) to date for the critical point
search ~\cite{Adam:2020unf}. As we go from observables involving lower order moments
($S\sigma$) to higher order moments ($\kappa\sigma^2$),
deviations between central and peripheral collisions for the measured
values increases. Central collisions $\kappa \sigma^{2}$ data show a
non-monotonic variation with collision energy with respect to the
statistical baseline of $\kappa \sigma^{2}$ = 1 at a significance of
$\sim$ 3$\sigma$~\cite{Adam:2020unf}. The deviations of
$\kappa \sigma^{2}$ below the baseline are
qualitatively  consistent with theoretical considerations including a
critical point~\cite{Stephanov:2011pb}.  In addition, experimental data
show deviation from heavy-ion collision models without a critical
point.  This can be seen from the table~\ref{tab1} which shows values of a $\chi^2$ test between
the experimental data and various models. In all cases, within 7.7 $<$
$\sqrt{s_{NN}}$ (GeV) $<$  27, the $\chi^{2}$ tests return $p$-values that
are less than 0.05. This implies that the monotonic energy dependence
from all of the models are statistically inconsistent with the data. 
Although a non-monotonic
variation of the experimental data with collision energy looks
promising for the QCD critical point search, a more robust conclusion
can be derived when the uncertainties get reduced and significance
above $5\sigma$ is reached.  This is the plan for the RHIC Beam Energy
Scan Phase-II program.

\begin{table}
	\caption{The $p$ values of a $\chi^2$ test between data and
          various models for the  $\sqrt{s_{NN}}$ dependence of $\it{S}\sigma$  and
          $\kappa\sigma^{2}$  values of net-proton
          distributions in 0-5\% central Au+Au collisions. The results
          are for the  $\sqrt{s_{NN}}$ range 7.7 to 27 GeV
          ~\cite{Adam:2020unf} which is the relevant region for the
          physics analysis presented here.}
	\centering   
	\begin{tabular}{|c|c|c|c|c|}
		\hline	
		Moments & HRG GCE & HRG EV & HRG CE &
                                                                  UrQMD  \\
          	 &  & (r = 0.5 fm)& &  \\
		\hline 
 $\it{S}\sigma$ & $<$ 0.001 &  $<$ 0.001 & 0.0754 & $<$ 0.001 \\
\hline 
$\kappa\sigma^{2}$ & 0.00553& 0.0145  & 0.0450 & 0.0221\\
\hline 
	\end{tabular}
	\label{tab1}
\end{table}

\subsection{Comparison to Lattice QCD inspired fits}
\begin{figure}[htb]
\centerline{%
\includegraphics[width=8.5cm]{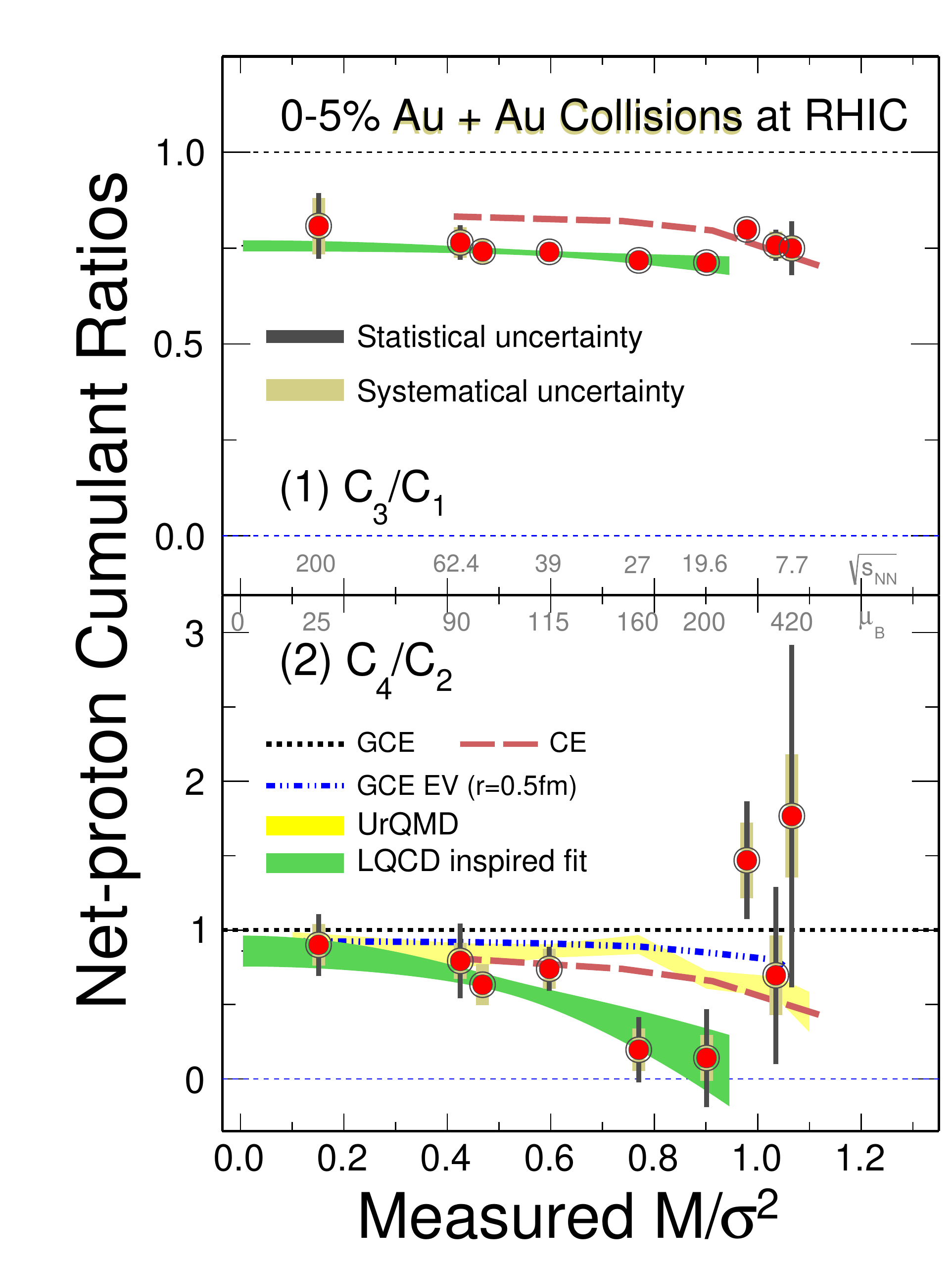}}
\caption{Net-proton cumulant ratios as a function of
  $M/\sigma^{2}$. Also shown are the expectations from different
  variants of HRG model (lines), UrQMD (yellow band) and LQCD inspired fits (green bands)~\cite{Bazavov:2017tot}.}
\label{Fig:F4}
\end{figure}

\begin{figure}[htb]
\centerline{%
\includegraphics[width=8.5cm]{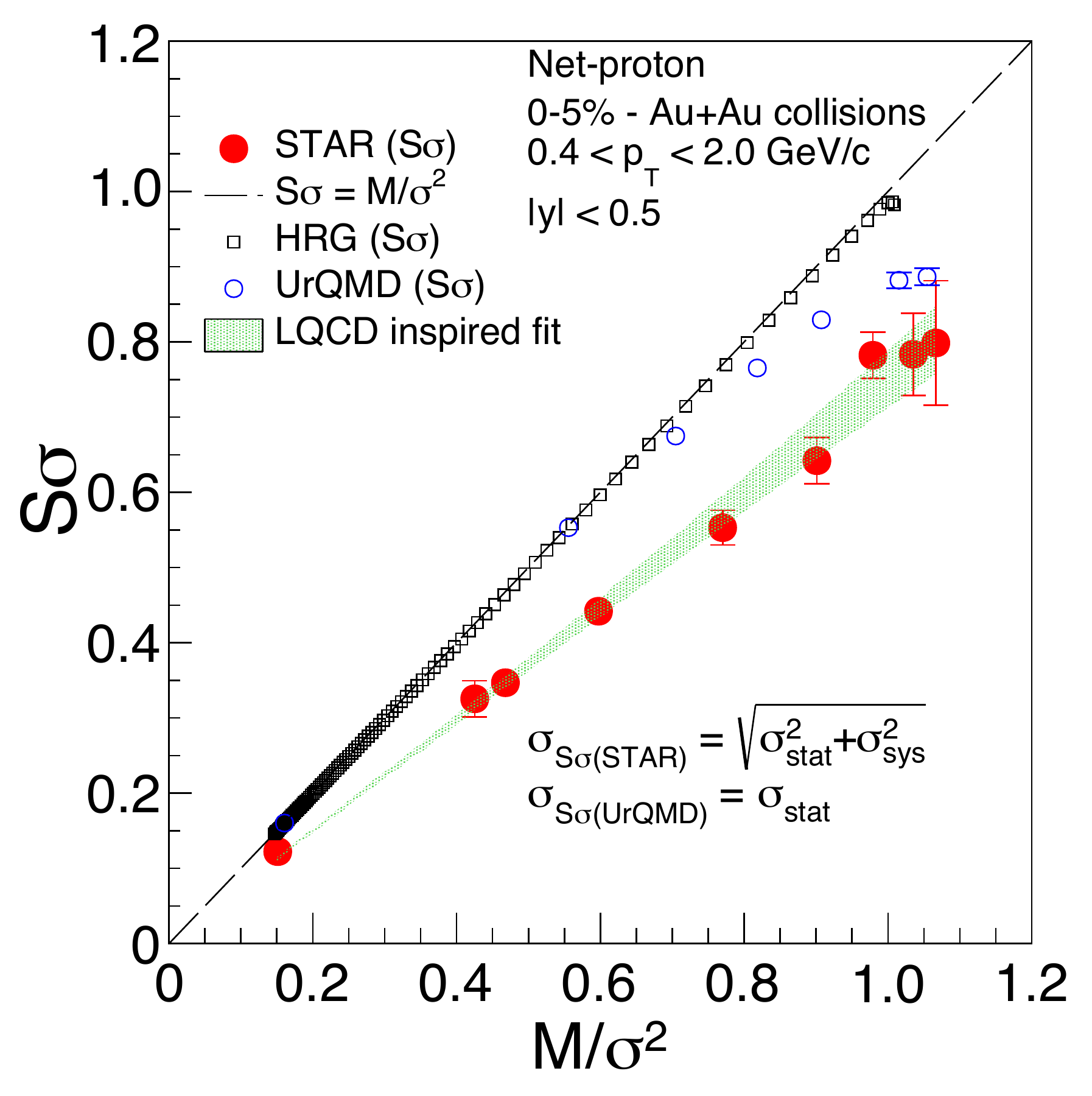}}
\caption{$S\sigma$ versus the $M/\sigma^{2}$ of net-proton distribution
  in high energy heavy-ion collisions. Also shown are the expectation
  from HRG, UrQMD and LQCD inspired fits~\cite{Bazavov:2017tot}.}
\label{Fig:F5}
\end{figure}

In the previous sub-section we have seen that the data deviates from the
expectations based on UrQMD and HRG models. Figures~\ref{Fig:F4} and
~\ref{Fig:F5} show that several features of the data are
qualitatively consistent with LQCD calculations of net baryon-number fluctuations
up to NLO in $\mu_B/T$~\cite{Bazavov:2017tot}. Specifically, (a) $M/\sigma^2 >
S\sigma$, where $M$ is the mean of the net-proton distribution; $C_{3}/C_{1}$ is smaller than unity and tending to
decrease with increasing $M/\sigma^2$; and
with increasing $M/\sigma^2$,  the cumulant ratio $C_{4}/C_{2}$ departs further away from unity 
than the ratio $C_{3}/C_{1}$ for $\sqrt{s_{_{NN}}}\ge 19.6$~GeV. The
LQCD inspired fits are of the form: $C_{3}/C_{1}$ = $p_{0}$ +
$p_{1}$ $(C_{1}/C_{2})^{2}$; $C_{4}/C_{2}$ = $p_{2}$ +
$p_{3}$ $(C_{1}/C_{2})^{2}$ and $C_{3}/C_{2}$ = $p_{0}$ $C_{1}/C_{2}$+
$p_{1}$ $(C_{1}/C_{2})^{3}$. Where $p_{0}$, $p_{1}$, $p_{2}$, and
$p_{3}$ are fit parameters and we have used the equivalence between
product of the moments and ratios of cumulants as $C_{1}/C_{2}$ =
$M/\sigma^2$; $C_{3}/C_{1}$ = $S\sigma^3/M$ and $C_{4}/C_{2}$ =
$\kappa\sigma^2$.  The good agreement between data and LQCD inspired
fits for $\sqrt{s_{NN}}$ range between 200 to 19.6 GeV, suggests that the heavy-ion collisions have produced a strongly
interacting QCD matter.

\section{Experimental programs for high baryon density}

\begin{figure}[htb]
\centerline{%
\includegraphics[width=9.0cm]{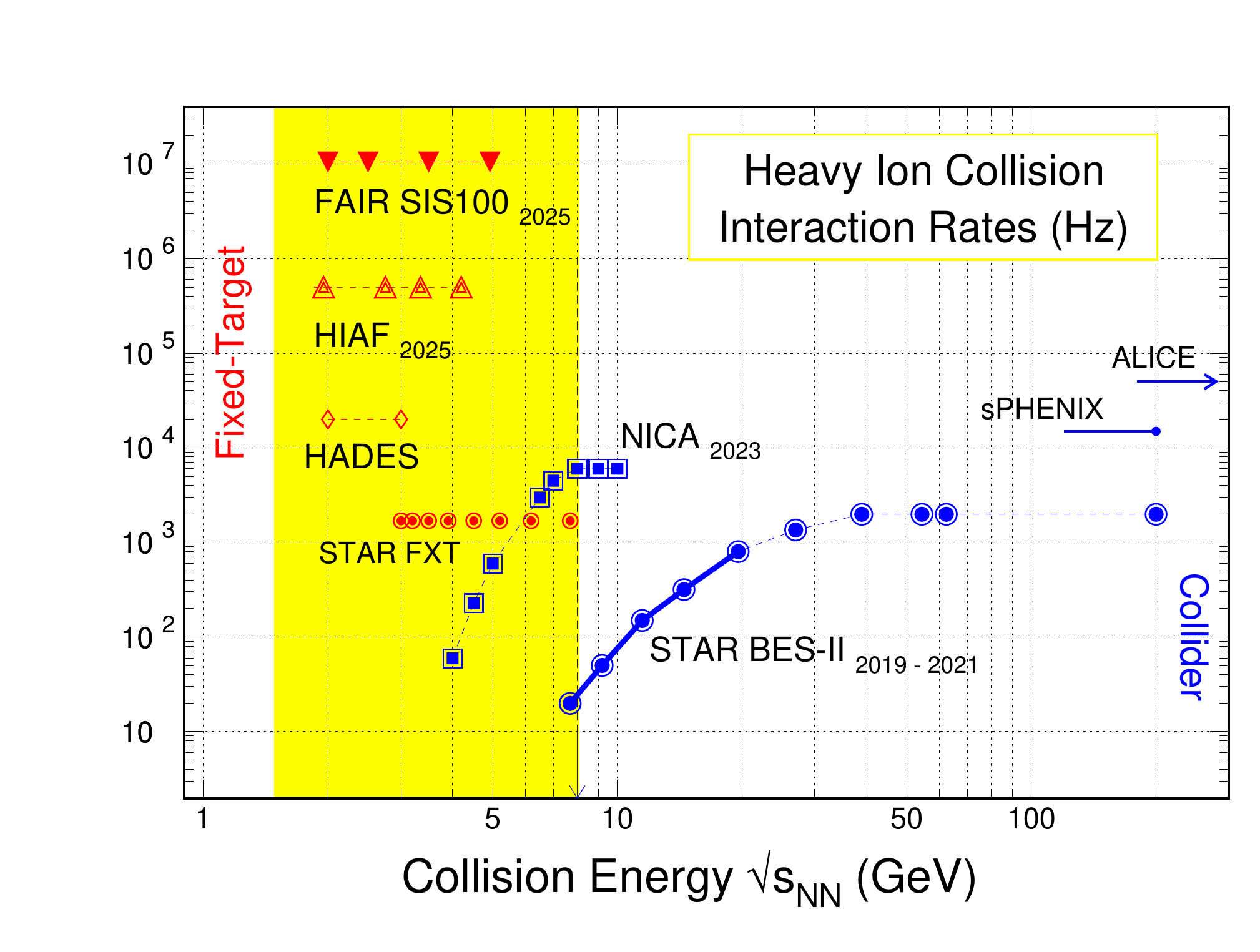}}
\caption{Interaction rates (in Hz) for high-energy nuclear collision
  facilities as a function of $\sqrt{s_{NN}}$~\cite{Fukushima:2020yzx}. Accelerators in
  collider mode are shown by blue symbols (ALICE, sPHENIX, RHIC BES-II
  and NICA) and those operating in fixed
  target mode by red symbols (STAR fixed traget (FXT), FAIR (CBM,
  SIS), HADES, and HIAF).}
\label{Fig:F6}
\end{figure}

As seen from the measurements discussed in previous section, to
complete the critical point search program a high statistics phase -
II of the beam energy scan program at RHIC is needed. In addition,
future new experiments, which are all designed with high rates, large
acceptance, and the state-of-the-art particle identification,
at the energy region where baryon density is high, i.e.,
500 MeV $< \mu_{B} <$ 800 MeV, see Fig.~\ref{Fig:F6}, will be needed.
The new facilities for studying high baryon density matter includes
(a) Nuclotron-based Ion Collider fAcility (NICA) at the Joint Institute
for Nuclear Research (JINR), Dubna, Russia~\cite{Geraksiev:2019fon}, (b) Compressed Baryonic Matter (CBM) at Facility for  
Antiproton and Ion Research (FAIR), Darmstadt, Germany~\cite{Ablyazimov:2017guv},
and (c) CSR External-target Experiment (CEE) at High Intensity heavy-ion Accelerator Facility (HIAF),
Huizhou, China~\cite{Ruan:2018fpo}.  

\section{Summary and Outlook}
The workshop dealt with two topics: Criticality and hadron resonance
gas models.

{\it Criticality:} A robust and vibrant research program is now established both
experimentally (several facilities)  and theoretically to study
the QCD phase structure~\cite{Gupta:2011wh} and seeking for the QCD critical point in the phase
diagram. The observables are well established and the results from a
first systematic measurements are promising. 

{\it Thermal models:} Another success story has been use of hadron
resonance gas models to extract freeze-out dynamics, provide evidences
for local thermalisation in heavy-ion collisions and act as baseline
for several measurements in heavy-ion collisions. This can be extended
further to test the details of the model, like GCE vs. CE, and
applications to higher order fluctuations to probe true thermal
nature of the system formed in heavy-ion collisions~\cite{Gupta:2020pjd}.

{\it High baryon density:} Gradual shift of attention of the heavy-ion
community is expected towards a return to the low energy collisions, where
state-of-art accelerator facility with large luminosity and much advances detector systems
with excellent particle identification will allow us to
unravel the physics of a rotating high baryon density QCD matter
subjected to magnetic field, similar to the neutron stars.
\\

{\it \b Acknowledgments} F. Karsch, V.~Koch, A. Pandav, and K.~Redlich for exciting
discussions. We also thank the colleagues from STAR and ALICE 
collaborations.  B.M. was supported in part by the Chinese Academy of Sciences  
President's International Fellowship Initiative
and J C Bose Fellowship from Department of Science of Technology, Government of India.
N.X. was supported in part by the Chinese NSF grant No.11927901 and the US DOE grant No.KB0201022.

\end{document}